\documentclass{elsart}

\usepackage{natbib}

\usepackage{epsfig}

\usepackage{amssymb}

\begin{document}

\begin{frontmatter}



\title{Particle Acceleration, Magnetic Field Generation,
and Emission in Relativistic Shocks}
\vspace*{-0.5cm}
\author[label1]{K.-I. Nishikawa,}
\author[label2]{P. Hardee,}
\author[label3]{C. B. Hededal,}
\author[label4]{G. Richardson,}
\author[label5]{R. Preece,}
\author[label6]{H. Sol,}
\author[label7]{and G. J. Fishman}
\address[label1]{National Space Science and Technology Center,
  Huntsville, AL 35805 USA}

\address[label2]{Department of Physics and Astronomy,
  The University of Alabama,
  Tuscaloosa, AL 35487 USA}

\address[label3]{Niels Bohr Institute, Department of Astrophysics,
Juliane Maries Vej30, 2100 K\o benhavn \O, Denmark}

\address[label4]{Department of Mechanical and Aerospace Engineering
University of Alabama in Huntsville Huntsville, AL 35899 USA}

\address[label5]{Department of Physics,
  University of Alabama in Huntsville,
  Huntsville, AL 35899 and National Space Science and Technology Center,
  Huntsville, AL 35805 USA}

\address[label6]{LUTH, Observatore de Paris-Meudon, 5 place Jules Jansen
92195
   Meudon Cedex, France}

\address[label7]{NASA-Marshall Space Flight Center, \\
National Space Science and Technology Center,
  Huntsville, AL 35805 USA}
\vspace*{-0.3cm}
\begin{abstract}
Shock acceleration is a ubiquitous phenomenon in astrophysical
plasmas.  Plasma waves and their associated instabilities (e.g.,
Buneman, Weibel and other two-stream instabilities) created in
collisionless shocks are responsible for particle (electron,
positron, and ion) acceleration. Using a 3-D relativistic
electromagnetic particle (REMP) code, we have investigated particle
acceleration associated with a relativistic  jet front propagating
into an ambient plasma. We
find small differences in the results for no ambient and modest
ambient magnetic fields. Simulations show that the Weibel
instability created in the collisionless shock front accelerates jet
and ambient particles both perpendicular and parallel to the jet
propagation direction. The small scale magnetic field structure
generated by the Weibel instability is appropriate to
the generation of ``jitter'' radiation from deflected electrons
(positrons) as opposed to synchrotron radiation. The jitter radiation
resulting from small scale magnetic field structures may be
important for understanding the complex time structure and spectral
evolution observed in gamma-ray bursts or other astrophysical
sources containing relativistic jets and relativistic collisionless
shocks.

\end{abstract}

\vspace*{-0.3cm}
\begin{keyword}
Relativistic shocks, Weibel instability, Particle acceleration,
Magnetic field generation, Radiation


\end{keyword}

\end{frontmatter}

\parindent 20pt

\vspace*{-1.3cm}
\section{Introduction}
\label{int}

\vspace*{-0.6cm}
Nonthermal radiation observed from astrophysical
systems containing relativistic jets and shocks, e.g., active
galactic nuclei (AGNs), gamma-ray bursts (GRBs), and Galactic
microquasar systems usually has power-law emission spectra. In most
of these systems, the emission is thought to be generated by
accelerated electrons through the synchrotron and/or inverse Compton
mechanisms. Radiation from these systems is observed in the radio
through the gamma-ray region. Radiation in optical and higher
frequencies typically requires particle re-acceleration in order to
counter radiative losses.

Particle-in-cell (PIC) simulations can shed light on the physical
mechanism of particle acceleration that occurs in the complicated
dynamics within relativistic shocks.  Recent PIC simulations using
injected relativistic electron-ion jets show that acceleration
occurs within the downstream jet, rather than by the scattering of
particles back and forth across the shock as in Fermi acceleration
(Frederiksen et al.\ 2003, 2004; Hededal et al. 2004; Nishikawa et
al.\ 2003, 2004, 2005). Silva et al.\ (2003) have presented
simulations of the collision of two inter-penetrating
electron-positron plasma shells as a model of an astrophysical
collisionless shock. In the electron-positron simulations performed
with counter-streaming jets (Silva et al.\ 2003), shock dynamics
involving the propagating jet head (where Fermi acceleration may
take place) was not investigated. In general, these independent
simulations have confirmed that relativistic jets excite the Weibel
instability (Weibel 1959).  The Weibel instability generates current
filaments and associated magnetic fields (Medvedev and Loeb 1999),
and accelerates
electrons (Silva et al.\ 2003; Frederiksen et al.\ 2003, 2004;
Nishikawa et al.\ 2003, Hededal et al. 2004).

In this paper we present new simulation results of particle
acceleration and magnetic field generation for relativistic
electron-positron and electron-ion shocks using 3-D relativistic
electromagnetic particle-in-cell (REMP) simulations.
In our new
simulations, electron-positron and electron-ion relativistic jets
are injected into electron-positron and electron-ion plasmas in order to
study the dynamics of a relativistic collisionless shock both with
and without an initial ambient magnetic field.

\vspace*{-0.7cm}
\section{Simulation Setup and results}

\vspace*{-0.5cm}
Four simulations were performed using an $85 \times
85 \times 320$ grid with a total of 180 million particles (27
particles$/$cell$/$species for the ambient plasma) and an electron
skin depth, $\lambda_{\rm ce} = c/\omega_{\rm pe} = 9.6\Delta$,
where $\omega_{\rm pe} = (4\pi e^{2}n_{\rm e}/m_{\rm e})^{1/2}$ is
the electron plasma frequency and $\Delta$ is the grid size
(Nishikawa et al. 2004).
%
In two other simulations an electron-positron jet is injected into a
magnetized and unmagnetized electron-positron ambient plasma and in two
simulations an electron-ion jet is injected into a magnetized and
unmagnetized electron-ion ambient plasma.  The choice of parameters and
simulations allows comparison with previous simulations (Silva et
al.\ 2003; Frederiksen et al.\ 2003, 2004; Hededal et al. 2004;
Nishikawa et al.\ 2003, 2004, 2005).

The electron number density of the jet is
$0.741n_{\rm b}$, where $n_{\rm b}$ is the density of ambient
(background) electrons. The average jet velocity $v_{\rm j} =
0.9798c$, and the Lorentz factor is 5 (2.5 MeV).
The jets are cold
($v^{\rm e}_{\rm j, th} = v^{\rm p}_{\rm j, th} = 0.01c$ and
$v^{\rm i}_{\rm j, th} = 0.0022c$) in the rest frame of the 
ambient palsma.
Electron-positron plasmas have mass ratio $m_{\rm p}/m_{\rm e} \equiv
m_{\rm e^+}/m_{\rm e^-} = 1$ and electron-ion plasmas have $m_{\rm
i}/m_{\rm e} = 20$.  The electron and ion thermal velocities in the
ambient plasmas are $v^{\rm e}_{\rm th} = 0.1c$ and $v^{\rm i}_{\rm th}
= 0.022c$, respectively, where $c$ is the speed of light.
The time step $\Delta t = 0.013/\omega_{\rm pe}$,
the ratio $\omega_{\rm pe}/\Omega_{\rm e} = 11.5$, and the Alfv\'en
speed (for electrons) $v_{\rm Ae} \equiv (\Omega_{\rm e}/\omega_{\rm
pe})c = 8.66\times 10^{-2}c$. With the speed of an Alfv\'en wave given by
$v_{\rm A} = [V_{\rm A}^{2}/(1 + V_{\rm A}^{2}/c^{2})]^{1/2}
= 6.10 \times 10^{-2}c$
where $V_{\rm A} \equiv [B^{2}/4\pi (n_{\rm e} m_{\rm e}
+ n_{\rm p} m_{\rm p})]^{1/2} = 6.12 \times 10^{-2}c$, the Alfv\'en Mach number
$M_{\rm A} \equiv v_{\rm j}/v_{\rm A} = 16.0$.
With a magnetosonic speed $v_{\rm ms} \equiv (v_{\rm
th}^{2} + v_{\rm A}^{2})^{1/2} = 0.132c$ the Magnetosonic Mach number $M_{\rm
ms} \equiv v_{\rm j}/v_{\rm ms} = 7.406$. At least approximately the
appropriate relativistic Mach numbers multiply these values by the
Lorentz factor.  Thus, in an MHD approximation we are dealing with a
high Mach number shock with $\gamma M >> 1$. The gyroradius of
ambient electrons and positrons with $v_{\perp} = v_{\rm th} = 0.1c$
is $11.1\Delta = 1.154\lambda_{\rm ce}$ (for ambient ions: $49.6\Delta =
5.16\lambda_{\rm ce}$).  All the Mach numbers with electron-ion jets
are approximately increased by $\sqrt{m_{\rm i}/m_{\rm e}} = \sqrt{20}
= 4.47$.


Current filaments resulting from development of the Weibel
instability behind the jet front are shown in Figs. 1a  and 1b at
time $t = 28.8/\omega_{\rm pe}$ for unmagnetized ambient plasmas. In
case (a) an electron-positron jet is injected into an
electron-positron ambient plasma. In case (b) an electron-ion jet is
injected into an electron-ion ambient plasma. The maximum values of
$J_{\rm y}$ are (a) 15.63 and (b) 10.7, respevtively. The
electron-positron jet shows larger amplitudes than the electron-ion
jet at the same simulation time and magnetic fields reduce the
maximum values, confirming previous simulation results (Nishikawa et
al. 2004). The effect of weak ambient magnetic fields affects the
growth rates of Weibel instability slightly as shown in Hededal and
Nishikawa (2004).

The heating and acceleration of jet electrons in directions parallel
and perpendicular to the flow is shown in Figure 1 for the
electron-positron case (1c and 1e) and for the electron-ion case (1d
and 1f).  The jet electrons are split into two parts: the injected
(blue: rear half ($Z< 160\Delta$) and shocked (red: front half ($Z>
160\Delta$). (The jet electrons are divided at $Z \sim 160\Delta$).)
Since in the case of electron-ion jet the Weibel instability grows
slightly the blue curves in Figs. 1d and 1f are considered as the
distributions of injected jet electrons. In both parallel and
perpendicular distributions, jet electrons are more accelerated in
the electron-positron case than in the electron-ion case.

%
\begin{figure}[ht]
\centerline{\psfig{figure=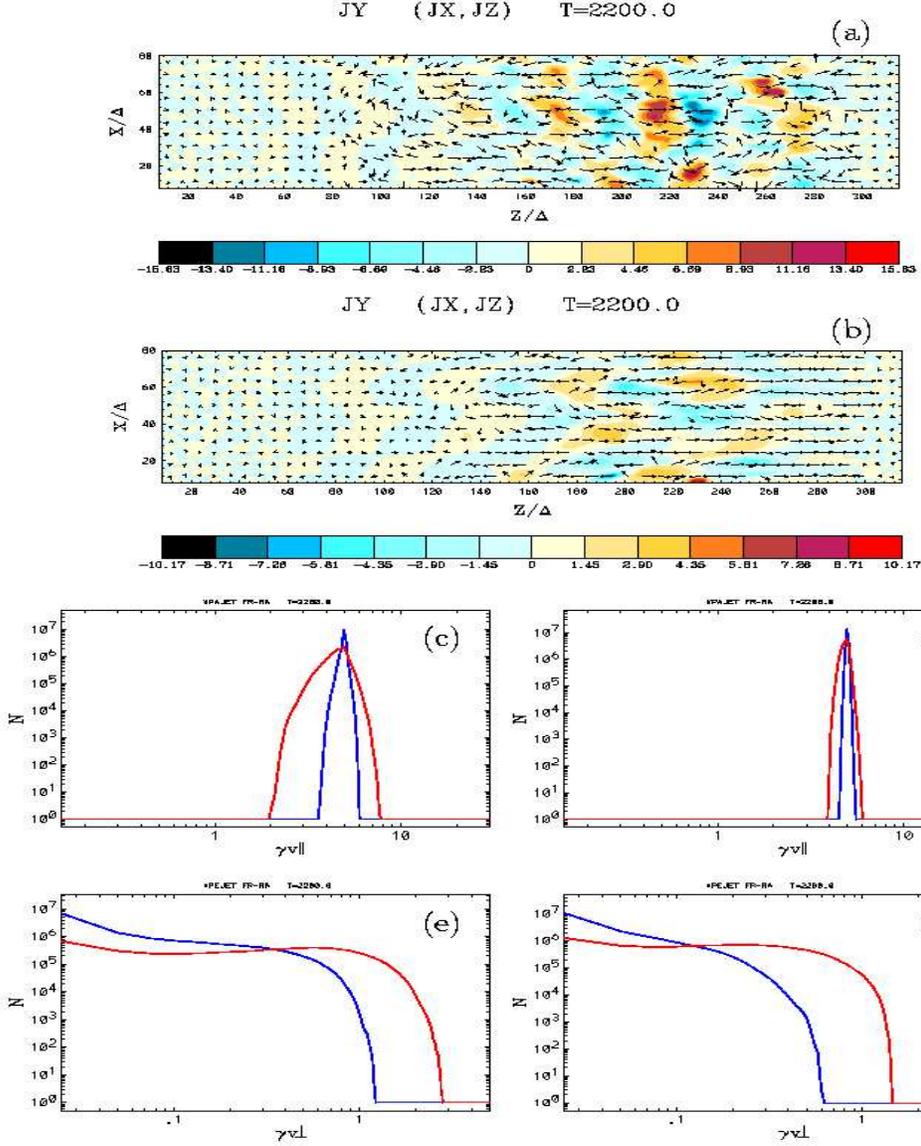,width=6.6in,height=6.5in}}
\vspace*{-1.0cm} \caption{Panels (a, c, e) and (b, d, f) refer to
the electron-positron and electron-ion cases, respectively
(unmagnetized cases). 2D images show the current density ($J_{\rm
y}$) at $t = 28.8/\omega_{\rm pe}$. Colors indicate the
$y$-component of the current density, $J_{\rm y}$ [peak: (a) 15.6,
(b) 10.7], and the arrows indicate $J_{\rm z}$ and $J_{\rm x}$.  The
injected (blue line) and shocked (red line) electron distributions
are shown as a function of $\gamma v_{\parallel}$ ((c) and (d)) and
$\gamma v_{\perp}$ (e and f) where $\gamma = (1 -(v^{2}_{\parallel}
+v^{2}_{\perp})/c^{2})^{-1/2}$.}

\end{figure}

The electrons are deflected by the transverse magnetic fields
($B_{\rm x}, B_{\rm y}$) via the Lorentz force: $-e({\bf v} \times
{\bf B})$, generated by current filaments ($J_{\rm z}$), which in
turn enhance the transverse magnetic fields (Weibel 1959; Medvedev
and Loeb 1999). The complicated filamented structures resulting from
the electron Weibel instability have diameters on the order of the electron
skin depth ($\lambda_{\rm ce} = 9.6\Delta$). This is in good
agreement with the prediction of $\lambda \approx
2^{1/4}c\gamma_{\rm th}^{1/2}/\omega_{\rm pe} \approx
1.188\lambda_{\rm ce} = 11.4\Delta$ (Medvedev and Loeb 1999). Here,
$\gamma_{\rm th} \sim 1$ is a thermal Lorentz factor. The filaments
are elongated along the direction of the electron-ion jets (b) (the
$z$-direction, horizontal in Figure 1). However, in the
electron-positron jets the current filaments have coalesced in the
transverse direction in the nonlinear stage. The transverse current
($J_{\rm x}$) in the electron-positron jets (a) shows significantly
more transverse variation than in the electron-ion jets (b).

 The acceleration of electrons has been reported in previous
work (Silva et al.\ 2003; Frederiksen et al.\ 2003, 2004; Nishikawa
et al.\ 2003, 2004, 2005; Hededal et al. 2004). We see that some of
the kinetic
energy (parallel velocity $v_{\parallel} \approx v_{\rm j}$) of the
jet electrons is transferred to the perpendicular velocity via the
electric and magnetic fields generated by the Weibel instability
as shown in Fig. 1. The strongest transverse and parallel acceleration 
of jet electrons accompanies
the strongest deceleration of electron flow and occurs between
$z/\Delta = 210 - 240$.  The transverse acceleration in the
electron-positron jets is over four times that in the electron-ion
simulations. The strongest acceleration takes place around the
maximum amplitude of perturbations due to the Weibel instability at
$z/\Delta \sim 220$ as seen in Figs. 1a and 1b.

\vspace*{-0.8cm}
\section{Summary and Discussion}

\vspace*{-0.5cm}
We have performed self-consistent,
three-dimensional relativistic particle simulations of relativistic
electron-positron and electron-ion jets propagating into magnetized
and unmagnetized electron-positron and electron-ion ambient plasmas.
The main acceleration of electrons takes place in the region behind 
the shock front. Processes in the relativistic collisionless shock are
dominated by structures produced by the Weibel instability.  This
instability is excited in the downstream region behind the jet head,
where electron density perturbations lead to the formation of
current filaments. The nonuniform electric field and magnetic field
structures associated with these current filaments decelerate the
jet electrons and positrons, while accelerating the ambient
electrons and positrons, and accelerating (heating) the jet and
ambient electrons and positrons in the transverse direction.

Other simulations with different skin depths and plasma frequencies
show that the growth rate and spatial structure of current filaments
generated by the Weibel instability scale with the plasma frequency
and the skin depth (Nishikawa et al. 2004).  An additional
simulation in which an electron-ion jet is injected into a ambient
plasma with perpendicular magnetic field shows magnetic reconnection
due to the generation of an antiparallel magnetic field generated by
bending of jet electron trajectories, consequently jet electrons are
subject to strong non-thermal acceleration (Hededal and Nishikawa 2004).

These simulation studies have provided new insights for particle
acceleration and magnetic field generation. Further research is
required to develop radiation models based on these microscopic
processes.

{\bf Acknowledgments:} K. Nishikawa is a NRC Senior Research Fellow
at NASA Marshall Space Flight Center. This research (K.N.) is
partially supported by the National Science Foundation awards
ATM-0100997, and INT-9981508. P. Hardee
acknowledges partial support by a National Space Science and
Technology (NSSTC/NASA) award.  The simulations have been performed
on ORIGIN 2000 and IBM p690 (Copper) at the National Center for
Supercomputing Applications (NCSA) which is supported by the
National Science Foundation.

\end{document}